\documentclass[aps,prd,preprint,tightenlines,nofootinbib,showpacs,byrevtex]{revtex4}
\usepackage{amssymb,latexsym}
\usepackage{amsmath,amsbsy,bbm}
\usepackage{epsfig,bm}
\usepackage{graphicx,comment}
\DeclareMathOperator{\tr}{tr}
\DeclareMathOperator{\Erfc}{Erfc}

\begin{document}
\def\a{{\alpha}}
\def\b{{\beta}}
\def\d{{\delta}}
\def\D{{\Delta}}
\def\e{{\varepsilon}}
\def\g{{\gamma}}
\def\G{{\Gamma}}
\def\k{{\kappa}}
\def\l{{\lambda}}
\def\L{{\Lambda}}
\def\m{{\mu}}
\def\n{{\nu}}
\def\o{{\omega}}
\def\O{{\Omega}}
\def\S{{\Sigma}}
\def\s{{\sigma}}
\def\th{{\theta}}

\def\ol#1{{\overline{#1}}}

\def\Dslash{D\hskip-0.65em /}
\def\Dtslash{\tilde{D} \hskip-0.65em /}

\def\CPT{{$\chi$PT}}
\def\QCPT{{Q$\chi$PT}}
\def\PQCPT{{PQ$\chi$PT}}
\def\tr{\text{tr}}
\def\str{\text{str}}
\def\diag{\text{diag}}
\def\order{{\mathcal O}}

\def\cC{{\mathcal C}}
\def\cB{{\mathcal B}}
\def\cT{{\mathcal T}}
\def\cQ{{\mathcal Q}}
\def\cL{{\mathcal L}}
\def\cO{{\mathcal O}}
\def\cA{{\mathcal A}}
\def\cQ{{\mathcal Q}}
\def\cR{{\mathcal R}}
\def\cH{{\mathcal H}}
\def\cW{{\mathcal W}}
\def\cM{{\mathcal M}}
\def\cD{{\mathcal D}}
\def\cN{{\mathcal N}}
\def\cP{{\mathcal P}}
\def\cK{{\mathcal K}}
\def\Qt{{\tilde{Q}}}
\def\Dt{{\tilde{D}}}
\def\St{{\tilde{\Sigma}}}
\def\cBt{{\tilde{\mathcal{B}}}}
\def\cDt{{\tilde{\mathcal{D}}}}
\def\cTt{{\tilde{\mathcal{T}}}}
\def\cMt{{\tilde{\mathcal{M}}}}
\def\At{{\tilde{A}}}
\def\cNt{{\tilde{\mathcal{N}}}}
\def\cOt{{\tilde{\mathcal{O}}}}
\def\cPt{{\tilde{\mathcal{P}}}}
\def\cI{{\mathcal{I}}}
\def\cJ{{\mathcal{J}}}

\def\eqref#1{{(\ref{#1})}}

\title{Flavor Twisted Boundary Conditions and Isovector Form Factors}

\author{ Brian C.~Tiburzi}
\email[]{bctiburz@phy.duke.edu}
\affiliation{Department of Physics,
Duke University,
Box 90305,
Durham, NC 27708-0305}

\date{\today}


\begin{abstract}
We use vector flavor symmetry to relate form factors of isospin changing operators to isovector form factors.
Flavor twisted boundary conditions in lattice QCD thus allow isovector form factors of twist-two operators, e.g, 
to be computed at continuous values of the momentum transfer. 
These twisted boundary conditions, moreover, are implemented only in the valence sector.
Effects of the finite volume must be addressed to extract isovector moments and radii at zero lattice momentum. 
As an example, we use chiral perturbation theory to assess the volume effects in extracting the isovector 
magnetic moment of the nucleon from simulations with twisted boundary conditions. 
\end{abstract}
\maketitle

\section{Introduction}

Numerically simulating quantum chromodynamics (QCD) on a spacetime lattice 
enables the study of hadrons from first principles. Hadronic properties
computed from lattice QCD, however, suffer from a number of artifacts
due to the approximations involved in solving the theory numerically.
These artifacts include volume effects, lattice spacing errors, the
use of unphysically large quark masses, and the partially quenched approximation. 
There has been considerable recent effort to address the reduction of these
systematic errors using effective field theory (EFT) methods, see, e.g.~\cite{Bernard:2002yk}. 
With improved numerical algorithms and enlarged computing resources, we are entering 
a period where lattice data in conjunction with EFTs
will provide physical predictions for QCD.

Another restriction in lattice simulations is the available momentum. 
With periodic boundary conditions, the momentum and hence momentum transfers
are quantized. On current dynamical lattices the lowest non-zero momentum mode
is $\sim 500 \, \texttt{MeV}$. Chiral EFTs
predict the momentum transfer dependence of form factors.  With current 
lattice sizes, however, the applicability of these theories to describe the momentum dependence is questionable. 
The restriction to quantized momentum transfer
leads to a further impediment: the extraction of radii and moments 
that are only accessible in the near-forward limit is severely limited 
without \emph{ad hoc} models for the momentum dependence.
For a lattice of uniform spatial size $L$, periodic boundary conditions
yield momentum modes $\bm{q} = 2 \pi \bm{n} / L$, for $\bm{n} \in \mathbb{Z}^3$. 
To reduce the momentum granularity one must increase the lattice volume and thus generate 
new gauge configurations---an extremely costly solution.\footnote{%
One might reason that continuum moment equations could be employed 
to deduce moments and radii, e.g., using $\bm{x} \times \bm{J}(x)$ to determine
magnetic dipole moments, or $\bm{x}^2 J_4(x)$ for charge radii. 
On the lattice, however, these operators do not 
circumvent the restriction to quantized momentum~\cite{Wilcox:2002zt}.
}

The use of periodic boundary conditions is one of simplicity not necessity.
The quark fields need to periodic only up to some transformation which is 
a symmetry of the action. Thus if $U$ is a symmetry of the action
and $U^\dagger U = 1$, we can choose a twisted boundary condition for the 
generic field $\phi$ of the form 
\begin{equation} \notag
\phi(x_i + L) = U \phi(x_i)
,\end{equation}
while maintaining the single valuedness of the action.
Twisted boundary conditions are by no means new~%
\cite{Gross:1982at,Roberge:1986mm,Wiese:1991ku,Luscher:1996sc,Bucarelli:1998mu,Guagnelli:2003hw,Kiskis:2002gr,Kiskis:2003rd,Kim:2002np,Kim:2003xt},
and there has been renewed interest in their utility at producing continuous hadron momentum~%
\cite{Bedaque:2004kc,deDivitiis:2004kq,Sachrajda:2004mi,Bedaque:2004ax,Tiburzi:2005hg,Flynn:2005in,Guadagnoli:2005be,Aarts:2006wt}.

Obtaining hadronic states at continuous values of momentum does not solve the problem of coarse grained sampling of form factors, 
as the momentum transfer will still be quantized.
In \cite{Tiburzi:2005hg} it was shown that matrix elements of flavor changing operators, however, 
can be accessed at continuous values of the momentum transfer $\bm{q}$, of the form
$\bm{q} = 2 \pi \bm{n} / L + \delta \bm{\theta}/L$, where $\delta \bm{\theta}$
is a continuous parameter and is equal to the difference of twist angles of the flavors changed. 
As such flavor-changing operators have no self-contractions, the difference of twists is implemented
only in the valence quark sector to produce momentum transfer. These so-called partially twisted 
boundary conditions eliminate the need to regenerate gauge configurations for each value of the twisting parameters,
and allow moments and radii of flavor changing operators to be extracted at zero lattice momentum (up to 
volume corrections~\cite{Sachrajda:2004mi}).
This procedure has been studied in numerical simulations: for extracting 
$f_\pi$~\cite{Flynn:2005in}, and for determining $K \to \pi$ 
matrix elements~\cite{Guadagnoli:2005be}.

In this work, we point out that the vector flavor symmetry $SU(2)_V$ of QCD relates matrix elements 
of isospin changing operators to matrix elements of isovector operators. Thus the form factors of the
latter can be accessed at continuous momentum transfer by calculating flavor changing matrix elements
on the lattice. This result holds for quark bilinear operators of arbitrary
spin and Lorentz structure. 
Our discussion is organized as follows. In Sect.~\ref{s:SU(2)_V}, we derive
the relations between matrix elements related by an isospin rotation. Next in Sect.~\ref{s:extension}, 
we detail how these relations can be utilized on the lattice with partially twisted
boundary conditions. The dynamical effects of the boundary conditions  
are discussed in Sect.~\ref{s:example}, 
where the volume effects for the nucleon isovector 
magnetic moment are presented. We end with a brief summary in Sect.~\ref{summy}.

\section{Vector Current Conservation} \label{s:SU(2)_V}

The Lagrangian of two flavor QCD,\footnote{%
In this Sect., we limit our discussion to the two flavor 
theory for simplicity. We will explain below how the result 
generalizes to the relevant subgroup of the 
partially quenched theory and analogous three-flavor theories.}
the quark part of which is
\begin{equation}
\mathcal{L} 
= \sum_{j,k=1}^2 \ol{Q}{}^{\hskip 0.2em j} \left(
 \Dslash + m_Q \right)_j^{\hskip 0.3em k} Q_k
,\end{equation}
with $Q = ( u, d )^T$ and $m = \diag (m_u, m_d)$, 
has an exact $SU(2)_V$ symmetry in the isospin limit, $m_u = m_d$,
that cannot be spontaneously broken.
We shall work exclusively in the isospin limit.
Let us denote the generators of $SU(2)_V$ by $T^a$.
Noether's theorem yields the current
\begin{equation}
J^{a}_\mu(x) = \ol Q (x) T^a \gamma_\mu Q(x) 
,\end{equation}  
with conserved charges
\begin{equation}
\cQ^a = \int d\bm{x} \, J^{a0} (x)
,\end{equation}
that are the generators of isospin rotations.

Now consider a quark bilinear operator $\mathcal{O}^a$
of the form
\begin{equation} \label{eq:bilinear}
\mathcal{O}^a (x) = \ol Q(x) \, T^a \, \Gamma \, Q(x)
.\end{equation}
The $\Gamma$ represents any Dirac matrix
and any Lorentz tensor. For example, the twist-two operator
\begin{equation}
A^a_{\mu \mu_1 \ldots \mu_n} (x) = \ol Q (x) \, T^a \, \gamma_5 \gamma_{\{ \mu}
D_{\mu_1} \cdots D_{\mu_n \}} Q (x), 
\end{equation}
is such an $\mathcal{O}^a(x)$, \emph{etc}. We leave $\Gamma$ unspecified because
only the flavor structure is relevant 
for our discussion. 
It is straightforward to show that 
\begin{equation}
[ \cQ^a, \mathcal{O}^b(x) ] = i \varepsilon^{abc} \mathcal{O}^c(x)
.\end{equation}
Defining the usual isospin raising and lowering operators
$T^\pm = T^1 \pm i T^2$, we have
\begin{equation} \label{eq:key}
\mathcal{O}^\pm (x) = \mp [ \cQ^\pm, \mathcal{O}^3(x)]
.\end{equation}
These relations enable us to relate isospin 
changing matrix elements to isovector ones.

As an example, consider the neutron to proton matrix element
of $\mathcal{O}^+$. We have
\begin{equation} \label{eq:protonkey}
\langle p \, | \, \mathcal{O}^+  | \, n \rangle 
= 
\langle p \, | \, \mathcal{O}^3  | \, p \rangle
- 
\langle n \, | \, \mathcal{O}^3  | \, n \rangle 
,\end{equation} 
by virtue of Eq.~\eqref{eq:key}. This result
can be rewritten in various forms using 
the isoscalar combination
\begin{equation}
\mathcal{O}^I(x) = \ol Q(x) \mathbbm{1} \, \Gamma \, Q(x)
,\end{equation}
and the fact that
\begin{equation}
\langle p \, | \, \mathcal{O}^I  | \, p \rangle
- 
\langle n \, | \, \mathcal{O}^I  | \, n \rangle
= 
0 
.\end{equation}
For example,
$\langle p \, | \, \mathcal{O}^+  | \, n \rangle 
= 2 \langle p \, | \, \mathcal{O}^3  | \, p \rangle
= \langle p \, | \, \ol u \, \Gamma \, u  | \, p \rangle - \langle n \, | \, \ol u \, \Gamma \, u  | \, n \rangle$.
There is a  particularly useful way to rewrite the above relation
in Eq.~\eqref{eq:protonkey} 
for the case of the electromagnetic current operator
\begin{equation}
J_\mu^{\text{em}} (x) = \ol Q(x) \, \cQ \, \gamma_\mu Q(x)
,\end{equation}
with $\cQ = \diag ( 2/3, - 1/3 )$. This form is
\begin{equation} \label{eq:protonEMkey}
\langle p \, | \, \ol u \, \gamma_\mu \, d  | \, n \rangle 
= 
\langle p \, | \,  J_\mu^{\text{em}}  | \, p \rangle
- 
\langle n \, | \,  J_\mu^{\text{em}}  | \, n \rangle 
.\end{equation}

Of course one is not limited to baryonic matrix elements. 
Below we shall largely highlight the calculation of nucleon electromagnetic form factors 
as an example of how to utilize twisted boundary conditions.
Our results, however, apply to the arbitrary 
quark bilinear operators in Eq.~\eqref{eq:bilinear},
using the relation in Eq.~\eqref{eq:key} between the desired states.

\section{Implementation on the Lattice} \label{s:extension}

The above operator relations can be utilized to study isovector form factors
at continuous values of the momentum transfer. In this Sect., 
we describe how to utilize partially twisted boundary conditions for this 
purpose. Here we focus on the kinematical effects; while in Sect.~\ref{s:example},
we take up the dynamical effects at finite volume.

To see that the relations implied by Eq.~\eqref{eq:key} 
[or for a specific example, the relation in Eq.~\eqref{eq:protonEMkey}] 
allow lattice matrix elements to be accessed  
at continuous momentum transfer,  
first observe that there are no operator self contractions, thus, 
the momentum injection occurs in the valence sector alone.
To separate the valence and sea sectors, we use the partially quenched Lagrangian
\begin{equation}
\cL = \sum_{j,k=1}^6 \ol{Q}{}^{\hskip 0.2em j} \left(
 \Dslash + m_Q \right)_j^{\hskip 0.3em k} Q_k
.\label{eq:pqqcdlag}
\end{equation}
The six quark fields transform in the fundamental representation of the
graded $SU(4|2)$ group and appear in the vector 
$Q = (u, d, j, l, \mathfrak{u}, \mathfrak{d})^{\text{T}}$ that, 
in addition to the $u$ and $d$ quarks, has ghost quarks
$\mathfrak{u}$ and $\mathfrak{d}$, which cancel the closed valence loops, and two 
sea quarks $j$ and $l$. 
In the isospin limit, the quark mass matrix of $SU(4|2)$ reads
$m_Q = \diag(m_u, m_u, m_j, m_j, m_u, m_u)$.

In a finite box, the quark fields must satisfy boundary conditions
that preserve the single valuedness of the action. Such a choice
is afforded by twisted boundary conditions of the form
\begin{equation}
Q(x + L \hat{\bm{e}}_r) = \exp \left( i \bm{\theta}^a \cdot \hat{\bm{e}}_r \, \ol T {}^a_C \right) Q(x)  
,\end{equation}
where $\hat{\bm{e}}_r$ is a unit vector in the $r^{\text{th}}$ spatial direction and 
the block diagonal form of the supermatrices $\ol T {}^a_C$ is
\begin{equation} \label{eq:qtwist}
\ol T {}^a_C 
= 
\diag 
\left( T^a_C, 0, T^a_C \right)
.\end{equation} 
Here 
$T^a_C$
are the elements of the Cartan subalgebra of 
$U(2)$.\footnote{%
Any generator of the $U(2)$ algebra can actually be chosen for the twists because 
we work in the isospin limit. In this case, the boundary can change $u$-quarks 
into $d$-quarks and vice versa. Electric charge conservation requires that the electromagnetic current 
couple to the background field 
$B_\mu$. 
Additionally the valence quark propagators 
become flavor non-diagonal, but these flavor rotations could be used to inject momentum transfer. 
For ease of applicability in lattice simulations, however, we have chosen to work in the 
Cartan subalgebra, where flavors do not rotate at the boundary, but isospin symmetry
implies operator relations between flavor diagonal and flavor rotated currents.
}
As a consequence of Eq.~\eqref{eq:qtwist}, the sea quarks are periodic, and hence the twist angles 
$\bm{\theta}^a$ 
can be altered without generating new gauge configurations.

Defining new quark fields as 
$\Qt(x) = V^\dagger(x) Q(x)$, 
where 
$V(x) = \exp ( i \bm{\th}^a \cdot  \bm{x} \, \ol T {}^a_C / L )$, 
we can write the partially quenched Lagrangian as
\begin{equation} \label{eq:Ltwist}
\cL 
= 
\sum_{j,k=1}^6 \ol{\Qt}{}^{\hskip 0.2em j} 
\left(
\Dtslash + m_Q \right)_j^{\hskip 0.3em k} 
\Qt_k
,\end{equation}
where all 
$\Qt$ 
fields satisfy periodic boundary conditions, and the effect of twisting has
the form of a 
$U(1)$ 
gauge field:  
$\Dt_\mu = D_\mu + i B_\mu$, 
with 
$B_\mu = (0, \bm{\th}^a \, \ol T {}^a_C / L)$. 
For convenience we treat the twisting in the flavor basis:
$\bm{\th}^a \, \ol T {}^a_C = \diag (\bm{\th}^u, \bm{\th}^d, \bm{0}, \bm{0}, \bm{\th}^u, \bm{\th}^d  )$, 
and similarly for 
$B_\mu = \diag (B^u_\mu, B^d_\mu, 0, 0, B^u_\mu, B^d_\mu )$.
The constant field 
$B_\mu$ 
acts as flavor-dependent field momentum. 
Meson and baryon fields formed from these 
$\Qt$ 
quark fields also acquire flavor-dependent momentum via the background 
$U(1)$ 
field namely, 
\begin{equation} \label{eq:mmom}
\Dt_\mu \St = \partial_\mu \St + i [ B_\mu, \St ]
,\end{equation}
for the mesons~\cite{Sachrajda:2004mi}, and 
\begin{equation} \label{eq:bmom}
[\cDt_\mu \cBt (x)]^{ijk} = \partial_\mu \cBt^{ijk} (x) + i (B_\mu^i + B_\mu^j + B_\mu^k) \cBt^{ijk}(x)
,\end{equation}
for the baryons~\cite{Tiburzi:2005hg}.

In analogy with QCD, we assume that the infinite volume theory described by Eq.~\eqref{eq:Ltwist}
will undergo spontaneous symmetry breaking of the form 
$SU(4|2)_L \otimes SU(4|2)_R \to SU(4|2)_V$.
The remaining vector symmetry is explicitly broken by the quark mass difference $m_j \neq m_u$
from $SU(4|2)_V \to SU(2|2)_V \otimes SU(2)_V$. The valence generators of the graded $SU(2|2)_V$
symmetry again lead to the operator relations in Eq.~\eqref{eq:key}. Furthermore matrix element
relations, e.g. that in Eq.~\eqref{eq:protonEMkey}, remain valid because the external states 
contain only valence quarks.

\begin{figure}[tb]
  \centering
  \includegraphics[width=0.25\textwidth]{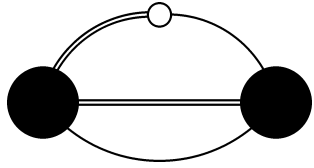}%
  \caption{Contraction encountered in the neutron-to-proton correlation function Eq.~\eqref{eq:corr}.
  The source and sink are denoted by filled circles, the operator insertion by an 
  open circle. Single lines represent the propagators of twisted $d$-quark fields, 
  while double lines represent the propagators of twisted $u$-quark fields. 
  }
  \label{f:latcor}
\end{figure}

Having related isovector matrix elements to isospin changing matrix elements in partially twisted QCD, we 
briefly recall how the latter can be determined at continuous momentum transfer 
from lattice correlators. 
To implement the twisted boundary conditions one uses interpolating fields
for the hadrons built of the $\Qt$ fields that are periodic but coupled to the background field $B_\mu$. 
For example, let us denote neutron and proton interpolating fields constructed in this way 
as $\tilde{\cN} (\bm{x},t)$ and $\tilde{\cP}(\bm{x},t)$. 
One then calculates the correlation function 
\begin{equation} \label{eq:corr}
C(t,t') 
= 
\sum_{\bm{x}, \bm{x'}} 
e^{-i \bm{P}\cdot \bm{x'}} 
\langle 0 | \cPt(\bm{x},t) \cOt^+ (\bm{x'},t') \ol \cNt (\bm{0},0) | 0 \rangle
,\end{equation}
where $\cOt^+$ is the valence isospin raising operator
\begin{equation}
\cOt^+ (\bm{x},t) = \overline{\tilde{u}} (\bm{x},t) \,  \tilde{\Gamma} \, \tilde{d} (\bm{x},t)
,\end{equation}
and the Dirac and Lorentz structure $\tilde{\Gamma}$ is arbitrary. The tilde
represents that any derivatives $D_\mu$ in $\Gamma$ appear as $\Dt_\mu$ in $\tilde{\Gamma}$.
A typical contraction contributing to this correlation function is depicted in Fig.~\ref{f:latcor}. 
The kinematic effect due to twisting is uncovered by expressing the correlation
function calculated on the lattice in terms of operators $\cN$ and $\cP$ built from the twisted quark fields $Q$. 
We have
\begin{eqnarray}
C(t,t') &=& 
\langle \cPt (\bm{0}, t) \cOt_\mu(t') \ol \cNt (\bm{P},0) \rangle 
\notag \\
&=& 
\langle \cP (\bm{B}_\cP, t) \cO_\mu(t') \ol \cN (\bm{P} + \bm{B}_{\cN},0) \rangle 
,\end{eqnarray}
where $\bm{P} = 2 \pi \bm{n} / L$ is the lattice momentum of the neutron, 
$\bm{B}_\cN = ( \bm{\th}^u + 2 \bm{\th}^d ) / L$, and  $\bm{B}_\cP = ( 2 \bm{\th}^u + \bm{\th}^d ) / L$
are the field momenta of the neutron and proton, respectively.
Notice the momentum transfer $\bm{q} = (\bm{\th}^u - \bm{\th}^d - 2 \pi \bm{n}) / L $ can be varied continuously.

Some final points are in order. 
Firstly, isoscalar quantities (which are notoriously difficult to calculate on the lattice)
of course cannot be deduced from these techniques. 
Next, the results trivially extend to the $SU(3)$ and $SU(6|3)$ flavor groups
because they contain the valence $SU(2)_V$ isospin subgroup. There are additional
matrix element relations in the limit of an exact valence $SU(3)_V$ 
symmetry; but because this symmetry is badly broken by the light-quark mass difference, 
we shall not investigate these relations. 
Finally, at finite volume, the presence of the $B_\mu$ field in the Lagrangian
breaks the $SU(2|2)_V$ symmetry and leads to modification
of our results. These modifications can be addressed systematically
in chiral perturbation theory (\CPT).

\section{Nucleon Isovector Magnetic Moment} \label{s:example}

In this Sect., we consider specifically the isovector magnetic form factor
of the nucleon. We first show the limitations in extracting the magnetic 
moment from lattice data at the smallest available lattice momentum transfer.
To show this is remedied by twisted boundary conditions,  
we calculate the isovector magnetic form factor
in heavy-baryon \CPT\ at finite volume
with partially twisted boundary conditions.

\subsection{Momentum Extrapolation} \label{s:momext}

In principle the momentum dependence of the nucleon form factors predicted
from heavy baryon \CPT%
~\cite{Jenkins:1990jv,Jenkins:1991es,Bernard:1992qa,Bernard:1995dp}
can be used to extrapolate lattice data 
down to zero momentum transfer in a model-independent way.
In \CPT, the isovector magnetic form factor
has the one-loop behavior~\cite{Bernard:1992qa,Jenkins:1993pi,Bernard:1995dp,Bernard:1998gv}
\begin{equation} \label{eq:ffinfvol}
F_2(q^2) = 
2 \mu_1 
- 
\frac{g_A^2 M}{2 \pi f^2} \int_0^1 dx \, m_\pi P_\pi(x,q^2)
-
\frac{g_{\D N}^2 M}{9 \pi^2 f^2}
\int_0^1 dx \, F[m_\pi P_\pi(x,q^2), \D]
,\end{equation}
where 
$P_\pi(x,q^2) = \sqrt{ 1 + x (1-x) q^2 / m_\pi^2}$, 
and the non-analytic function 
$F(m,\d)$ 
is given by
\begin{equation}
F(m,\d) 
= 
- \d \log \frac{m^2}{4 \d^2} 
+ 
\sqrt{\d^2 - m^2} \log 
\left( 
\frac{\d - \sqrt{\d^2 - m^2} + i \varepsilon}{\d + \sqrt{\d^2 - m^2}+ i \varepsilon}
\right)
.\end{equation}

\begin{figure}[tb]
  \centering
  \includegraphics[width=0.5\textwidth]{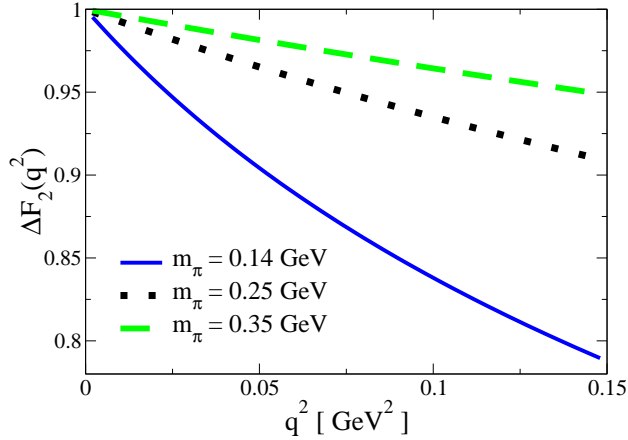}%
  \caption{Plot of the isovector magnetic form factor's deviation from linear $q^2$ behavior. 
  }
  \label{f:mom}
\end{figure}

Using estimates of the low-energy constants: $g_A = 1.25, |g_{\D N}| = 1.5$, $M = 0.94 \, \texttt{GeV}$,
$\Delta = 0.29 \, \texttt{GeV}$,  and $\mu_1 = 3.38$; we plot the deviation from linearity 
of the one-loop result for the $F_2$ form factor in Fig.~\ref{f:mom}
for various values of the pion mass. This is done using the function $\D F_2(q^2)$ defined by
\begin{equation}
\D F_2(q^2) = \frac{F_2(q^2) - F_2(0)}{q^2 F'_2(0)}
.\end{equation}
The plot shows considerable deviation from linearity at the physical pion mass.
For lattice pion masses around $0.25$ to $0.35 \, \texttt{GeV}$, there is 
a $\sim 5 - 10 \%$ deviation from linear behavior. As the lattice pion masses
are brought down at fixed lattice spacing (so that the minimal $q^2$ remains fixed), 
there is a clear trend toward non-linear behavior in $q^2$. 

We must keep in mind 
that the results plotted in the Figure receive $\sim 30 \%$ corrections for 
pion masses around $0.35 \, \texttt{GeV}$ due to higher-order terms in the chiral expansion
that scale as $m_\pi / \Lambda_\chi$~\cite{Meissner:1997hn,Puglia:1999th}. 
Such corrections are independent of $q^2$.
Beyond $q^2 \sim 0.1 \, \texttt{GeV} \, {}^2$ there are $\sim 30 \%$ corrections
from recoil $q^2 / M^2$ terms in loop graphs that contribute 
to the deviation from linearity~\cite{Kubis:2000aa}. 
The Figure shows that non-linear $q^2$ behavior is to be expected from the
magnetic form factor for moderate pion masses. The EFT, however, 
is at a loss to describe the lattice data at the minimal lattice momentum transfer
available on current dynamical lattices: $q^2 = 0.25 \, \texttt{GeV} \, {}^2$ lies
far off the axis in the plot, certainly too far to utilize the EFT.

\subsection{Twisted Boundary Conditions and Finite Volume Modifications} \label{s:finvol}

From our discussion in Sect.~\ref{s:extension}, we know that the
restriction to discrete lattice momentum can be circumvented by 
using partially twisted boundary conditions and calculating matrix elements
of the isospin raising operator, see Eq.~\eqref{eq:protonEMkey}. 
We demonstrate that this is the case by calculating the isovector
magnetic form factor using partially twisted baryon \CPT. Additionally by using this theory 
at finite volume, we can deduce the dynamical effects due to the boundary
conditions. These effects are systematic errors that must be removed to determine 
the form factor.

\subsubsection{Partially Twisted and Partially Quenched \CPT}

In the meson sector of partially quenched \CPT\ (\PQCPT)%
~\cite{Bernard:1994sv,Sharpe:1997by,Golterman:1998st,Sharpe:2000bc,Sharpe:2001fh}, 
the coset field 
$\Sigma$, 
which satisfies twisted boundary conditions, can be traded in
for the field $\St$ defined by
$\St (x) = V^\dagger(x) \S (x) V(x)$,
which is periodic at the boundary \cite{Sachrajda:2004mi}. 
In terms of this field, the Lagrangian of \PQCPT\ appears as
\begin{equation}
\cL = 
\frac{f^2}{8} \str \left( \Dt_\mu \St \Dt_\mu \St^\dagger \right)
- 
\l \, \str \left( m_Q^\dagger \St + \St^\dagger m_Q \right)
.\end{equation}
The action of the covariant derivative $\Dt^\mu$ is specified in Eq.~\eqref{eq:mmom}.

To include baryons into \PQCPT, one uses rank three flavor tensors \cite{Labrenz:1996jy}.
The spin-$\frac{1}{2}$ baryons are described by the $\bm{70}$-dimensional supermultiplet $\cB^{ijk}$, 
while the spin-$\frac{3}{2}$ baryons are described by the $\bm{44}$-dimensional supermultiplet $\cT_\mu^{ijk}$ %
\cite{Beane:2002vq}. 
The baryon flavor tensors we use, however, are twisted at the boundary of the lattice. 
Thus we define new tensors $\cBt^{ijk}$ and $\cTt_\mu^{ijk}$ both having the form~\cite{Tiburzi:2005hg}
\begin{equation}
\cBt_{ijk}(x) = V^\dagger_{ii}(x) V^\dagger_{jj}(x) V^\dagger_{kk}(x) \cB_{ijk}(x)
.\end{equation}
These baryon fields satisfy periodic boundary conditions and their free and leading-order interaction Lagrangian has 
been given in~\cite{Tiburzi:2005hg}.

In partially quenched QCD, the isovector vector current is defined by 
$J_\mu^{a}(x) = \ol Q(x)  \, \ol T {}^a  \gamma_\mu  Q(x)$. 
The choice of supermatrices 
$\ol T {}^a$ 
is not unique \cite{Golterman:2001qj}, 
even when one imposes the condition 
$\str \, \ol T {}^a = 0$.  
One should choose a form of the supermatrices that maintains the cancellation of valence and ghost quark loops
with an operator insertion~\cite{Tiburzi:2004mv,Detmold:2005pt}. 
For the flavor changing contributions we consider below, however, 
these operator self-contractions automatically vanish.  
For our calculation we require the action of 
$J_\mu^{a}$ 
in only the valence sector, and specify
the upper 
$2 \times 2$ 
block of 
$\ol T {}^{a}$ 
to be the usual isospin generators  
$T^a$. 
Henceforth we restrict our attention to the operator 
$J_\mu^{+} \equiv J_\mu^{1} + i J_\mu^{2}$.

In calculating the isovector magnetic moment of the nucleon at next-to-leading
order in the chiral expansion, there are local operators that contribute at tree level.
In \CPT, the leading isovector current operator is 
\begin{equation}
\d J_\mu^+ 
= 
\frac{\mu_1}{M} \partial_\nu 
\left( 
\ol N \sigma_{\mu \nu} T^+ N
\right)
.\end{equation}
In partially twisted \PQCPT, there are two terms in the leading isovector current operator 
\begin{equation}
\d J_\mu^+ 
= 
\frac{1}{M} \Dt_\nu 
\left[ 
\mu_\a \left(\ol \cBt \, \sigma_{\mu \nu} \cBt \, \ol T {}^+  \right)  
+  
\mu_\b \left(\ol \cBt \, \sigma_{\mu \nu}  \ol T {}^+ \cBt  \right) 
\right]
.\end{equation}
The contribution of these operators at tree level is proportional to the linear combination 
$\frac{1}{3} \mu_\a - \frac{1}{6} \mu_\b$, which is identical to 
the \CPT\ low-energy constant $\mu_1$ as can be demonstrated by matching~\cite{Beane:2002vq}.

\subsubsection{Partially Twisted Isovector Magnetic Form Factor}

\begin{figure}[tb]
  \centering
  \includegraphics[width=0.35\textwidth]{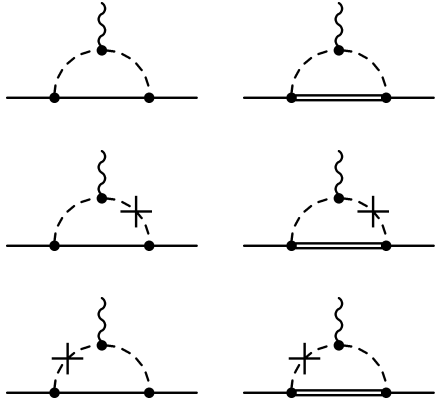}%
  \caption{One-loop contributions to the isospin transition matrix elements
  in partially twisted heavy baryon \CPT. Nucleons (deltas)
  are represented by single (double) lines, while mesons are represented by 
  dashed lines. The cross is the hairpin interaction, and the wiggly line
  shows the insertion of the isospin raising operator. 
  }
  \label{f:Nvecff}
\end{figure}

In the infinite volume limit, the isovector magnetic moment can be extracted from the
matrix element
\begin{equation}
\langle p(\bm{q}) \downarrow | J_3^+  | n(\bm{0}) \uparrow \rangle 
= 
\frac{- i q  }{2 M} F_2(q^2)
,\end{equation}
in the case where $\bm{q} = (0,q,0)$.
On the lattice, we can take both the source and sink to be at zero 
momentum, so that $\bm{q} = \bm{0}$. Momentum transfer can then be induced 
by giving the valence up and down quarks different twist angles. 
For simplicity we choose $\bm{B}^d = \bm{0}$ and $\bm{B}^u = (0,B,0)$.
Calculation of the infinite volume isovector form factor
then proceeds similarly to that above in Sect.~\ref{s:momext}. 
In partially twisted baryon \CPT,
there are additional diagrams that contribute involving the 
hairpin interaction. These diagrams are shown in Fig.~\ref{f:Nvecff}. 
The sum of all hairpin diagrams, however, vanishes.
With $m_j = m_u$,\footnote{%
When $m_j \neq m_u$, the results are identical to the partially 
quenched version of $F_2(q^2)$, the form of which can be inferred from expressions in~\cite{Beane:2002vq}. 
} 
the infinite volume contributions 
from the diagrams in the Figure are identical to $F_2(q^2)$ in Eq.~\eqref{eq:ffinfvol}
under the simple replacement $q \to B$. This is the kinematic effect 
we expect from raising isospin with a twisted $u$-quark.

Dynamical effects due to twisted boundary conditions 
arise from the propagation of the light Goldstone modes
to the boundary. The sensitivity of these modes to the boundary conditions
must be taken into account and can be done so in a model-independent way 
using baryon \CPT\ in finite volume. The modification 
to the effective theory is straightforward. In effect, we replace the integrals
over loop momenta with sums over the allowed\footnote{%
As is customary we treat the length of the time direction $T \gg L$, so that
for $k_\mu = (k_0, \bm{k})$, we can take $k_0$ to be continuous and $\bm{k}$ quantized
as above.}
modes $\bm{k} = 2 \pi \bm{n}/ L$.
The twisting is already taken into account in the effective theory by the
$U(1)$ gauge covariant derivative $\Dt_\mu$ above. The Poisson re-summation
formula then allows us to cast these sums into the infinite volume result 
plus the finite volume modification.

\begin{figure}[tb]
  \centering
  \includegraphics[width=0.65\textwidth]{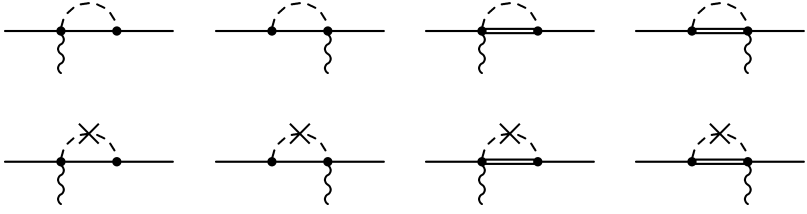}%
  \caption{Additional one-loop contributions to the isospin transition matrix elements.
  Diagram elements are the same as in Fig.~\ref{f:Nvecff}. These diagrams vanish for 
  periodic fields in a finite volume. 
  }
  \label{f:moreNvecff}
\end{figure}

There are, however, further contributions from using the partially twisted 
chiral theory in a box.\footnote{%
As with the pions~\cite{Sachrajda:2004mi},
the proton and neutron are no longer degenerate due to finite volume effects.
The volume induced isospin splittings are largest for small pion masses
and grow with $\th$. In a $2.5 \, \texttt{fm}$ box at $\th = \pi$, and 
at the physical pion mass,  
the splittings are $\sim 15 \%$ for the pions and $\sim 5 \%$ for the nucleons.
When the pion mass is twice as big, the splittings are $\sim 1 \%$ for both and are hence
neglected. 
}
Diagrams that ordinarily vanish in infinite volume
can now make contributions in a finite volume. This is the case for the diagrams
depicted in Fig.~\ref{f:moreNvecff}. Quite interestingly these diagrams 
are only non-vanishing in a finite volume with twisted boundary conditions.
When the twisting parameters vanish, so too does this finite volume effect.
This can easily be explained. In infinite volume the diagrams in Fig.~\ref{f:moreNvecff}
vanish due to $SO(4)$ rotational invariance, while in a periodic finite volume 
they vanish due to invariance under lattice rotations. Lattice rotational 
invariance is broken in the direction of the twisted boundary conditions, 
hence the diagrams make non-vanishing contributions.
Notice that the hairpin diagrams each vanish because the flavor-neutral 
mesons are additionally neutral under $B_\mu$.

Combining the infinite volume and finite volume results with
a twisted valence $u$-quark and specifying the case $m_j = m_u$, we arrive at
\begin{eqnarray} \label{eq:twistedanswer}
\langle p(\bm{0}) \downarrow | J_3^+  | n(\bm{0}) \uparrow \rangle 
&=& 
\frac{- i B }{2 M} \Bigg(
F_2(B^2)
-
\frac{g_A^2 M}{ 4 \pi^2 f^2 B} \cK_2(m_\pi, B \hat{y}, 0)
-
\frac{g_{\D N}^2 M}{36 \pi^2 f^2 B}
\cK_2(m_\pi, B \hat{y}, \D)
\notag\\
&&
+ 
\frac{3 M }{4 \pi^2 f^2}
\int_0^1 dx \, 
\Big\{ 
g_A^2 \cL_{33}[m_\pi P_{\pi}(x,B^2), x B \hat{y},0]
\notag \\
&& \phantom{bigspacererer} + 
\frac{2}{9} g_{\D N}^2 
\cL_{33}[m_\pi P_{\pi}(x,B^2), x B \hat{y},\D]
\Big\}
\Bigg)
,\end{eqnarray}
where 
$F_2(B^2)$ 
is given in Eq.~\eqref{eq:ffinfvol}. 
The effects of the finite volume are encoded in the functions $
\cK_2(m,\bm{B},\D)$ 
and 
$\cL_{33}(m,\bm{B},\D)$,
which  are defined in the Appendix.
In the limit that 
$B \to 0$, 
the result above accordingly vanishes: $B$ functions as a momentum transfer which is 
necessary for the magnetic form factor to be visible. 
Taking the derivative 
\begin{equation}
\lim_{B \to 0} 
\frac{2 M i}{B} 
\langle p(\bm{0}) \downarrow | J_3^+  | n(\bm{0}) \uparrow \rangle 
\notag
,\end{equation}
we can extract the magnetic moment $F_2(0)$ up to additive volume corrections. The corrections at 
$B = 0$ 
involving the 
$\cL_{33}(m_\pi, \bm{0}, \D)$ 
function are identical to the finite volume results in~\cite{Beane:2004tw}.\footnote{%
For comparison, we have $\cL_{33}(m,\bm{B} = \bm{0},\D) = \frac{4}{9} \mathcal{Y}(\D)$ of Ref.~\cite{Beane:2004tw}.
} 
Those results, however, were derived under the assumption that a momentum extrapolation to zero had been performed,
or alternately that one had employed a background magnetic field. Devoid of these assumptions, 
the current insertion method produces a second finite volume correction to the magnetic moment involving
\begin{equation}
\frac{ \partial \cK_2(m_\pi, B \hat{y}, \D)}{\partial B} \Big|_{B= 0} 
\notag
.\end{equation}

\begin{figure}[tb]
  \centering
  \includegraphics[width=0.55\textwidth]{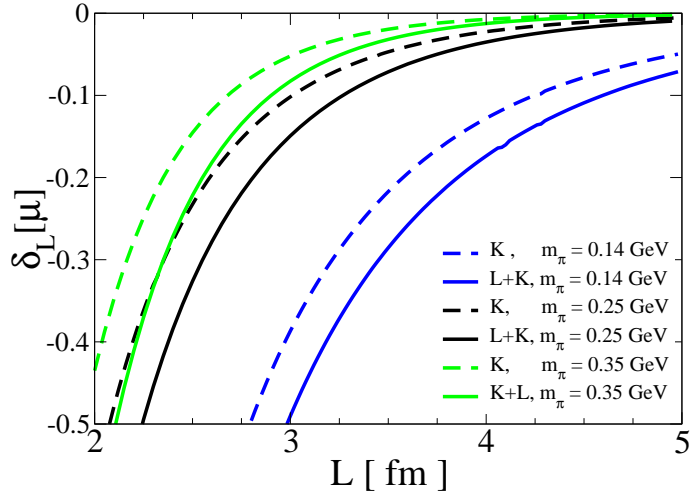}%
  \caption{Comparison of finite volume effects for the isovector magnetic moment.
Plotted versus $L$ is the relative difference $\d_L [\mu]$ for a few values of the 
pion mass. The total finite volume effect is denoted by $K + L$, while the contribution
to $\d_L [\mu]$ from $\partial \cK_2 / \partial B$ alone is denoted by $K$. 
  }
  \label{f:compare}
\end{figure}

\begin{figure}[tb]
  \centering
  \includegraphics[width=0.55\textwidth]{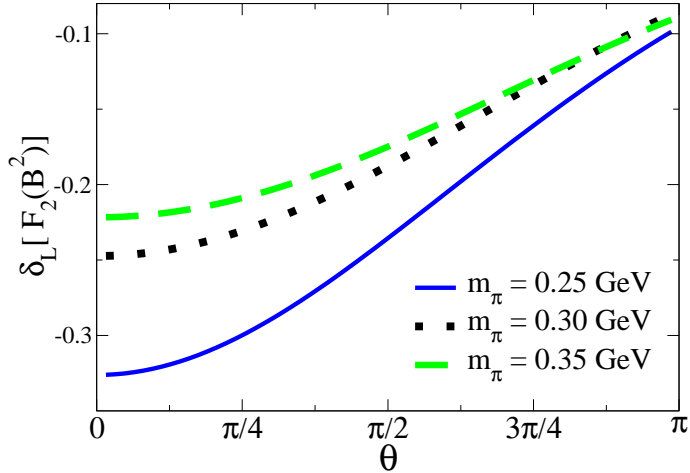}%
  \caption{Finite volume effects for extracting the isovector magnetic moment.
Plotted versus $\th$ is the relative difference $\d_L [F_2(B^2)]$ for a few values of the 
pion mass. The momentum transfer from twisting is $B = \th \times 0.079   \, \texttt{GeV}$ 
}
  \label{f:theta}
\end{figure}

To investigate the effect of the finite volume on the extraction of the isovector 
magnetic moment using twisted boundary conditions, we
define the relative difference 
\begin{equation} \label{eq:reldiff}
\d_L [F_2(B^2)] 
= 
\frac{\frac{2 M i }{B} \langle p(\bm{0}) \downarrow | J_3^+  | n(\bm{0}) \uparrow \rangle  -  F_2(B^2)}{F_2(B^2)}
.\end{equation} 
In the limit $B \to 0$, the difference is just $\d_L [F_2(0)] = \d_L [\mu]$, the relative 
difference in the magnetic moment. 
In Fig.~\ref{f:compare}, we plot $\d_L [\mu]$ as a function of $L$ for various
values of the pion mass to contrast our results with those of Ref.~\cite{Beane:2004tw}. 
We plot the total contribution to 
$\d_L[\mu]$ 
which arises from both the 
$\partial \cK_2 /\partial B$, 
and 
$\cL_{33}$ 
functions in Eq.~\eqref{eq:twistedanswer}, as well as just the contribution from 
$\partial \cK_2 /\partial B$. 
The latter is the dominant finite volume effect. 
Of course, to extract the magnetic moment, we require 
$\th \neq 0$
and thus we investigate 
$\d_L [F_2(B^2)]$ 
in Eq.~\eqref{eq:reldiff}. In Fig.~\ref{f:theta}, we fix the lattice volume at 
$2.5 \, \texttt{fm}$
and plot the relative difference 
$\d_L [F_2(B^2)]$ 
as a function of 
$\th$ 
for a few values of the pion mass.
We see that the effect of the finite volume decreases with 
$\th$. 
Generally the volume effects with momentum transfer (here the momentum transfer 
$q = B = \th / L$) 
are smaller than those at zero momentum transfer.\footnote{%
The actual behavior with respect to momentum transfer is damped oscillatory. 
The oscillations arise from the kinetic term $(\bm{k} + x \bm{q})^2$; 
but, in an $L = 2.5 \, \texttt{fm}$ box, 
they set in beyond the reach of the effective theory.
} 
In~\cite{Detmold:2005pt} this was anticipated due to the loop 
pion mass appearing as 
$m_\pi^2 + x(1-x) q^2 > m_\pi^2$. 
There is also a simple physical interpretation for this effect:
with a space-like momentum transfer the correlation function is
being probed on distances of order 
$\sim 1/ \sqrt{q^2}$. 
As 
$q^2$ 
increases, the resolving power of the virtual probe diminishes the volume effect.

While twisted boundary conditions
have introduced systematic error into the determination 
of the isovector moment, we stress that this is a controlled
error. In a fixed box size, we need only know the low-energy 
constants $g_A$, $g_{\D N}$, $\D$, and $f_\pi$ to remove this effect. 
Without twisting, one must rely on phenomenological or other fitting
functions which introduce uncontrolled error. 
With twisting, however, the lattice practitioner can approach the 
calculation of the isovector magnetic moment from various ways in conjunction
with the EFT.
For example, one can choose values of $\th \sim 0.1$ to eliminate the need
for a momentum extrapolation, \emph{cf} Fig.~\ref{f:mom}. One must then use
the EFT to remove the $\sim 25\%$ volume effects. Alternately one
could choose $\th \lesssim \pi$ to minimize the effect of the finite volume. 
For these values of $\th$ one then uses the momentum dependence predicted by 
the EFT, \emph{cf} Eq.~\eqref{eq:ffinfvol}, to obtain the magnetic moment.
Another virtue to the EFT approach is that 
higher-order corrections to the quark mass, volume, and momentum-transfer dependence
can be calculated to aid in the extrapolation.
Although we have focused on the magnetic moment, the expressions we have derived here
are also relevant for the isovector magnetic radius.

As a final point, an assumption inherent in our discussion is that no multi-particle 
thresholds are reached. For large enough momenta, multi-particle 
thresholds are inevitably reached. Volume corrections are no 
longer exponentially suppressed in asymptotic volumes; they become power law
and will likely dwarf any signal.
For the extraction of moments and radii of stable particles near zero momentum transfer, however, 
we are safely away from the multi-particle continuum.

\section{Summary} \label{summy}

We have related matrix elements of arbitrary quark bilinear operators that change isospin
to isovector combinations of those that do not change isospin. Using flavor twisted boundary 
conditions on the valence quark fields, form factors of these isovector matrix elements
can then be deduced from the lattice at continuous values of the momentum transfer. 
Isovector moments and radii can thus be determined from simulations at zero 
lattice momentum. 
Using twisted boundary conditions on the quark fields dramatically modifies the effect of the 
finite volume, even away from multi-particle cuts. 
This systematic effect can be handled with EFTs.
The determination of isovector moments and radii without any model-dependent assumptions about
the momentum dependence is thus well within reach of current resources.

\begin{acknowledgments}
We thank W.~Detmold for discussions,
and acknowledge the Institute for Nuclear Theory at the University of Washington 
for its hospitality and partial support during the course of this investigation.
This work is supported in part by the U.S.\ Dept.~of Energy,
Grant No.\ DE-FG02-05ER41368-0. 
\end{acknowledgments}

\appendix

\section*{Finite Volume Functions}

In this Appendix, we define and evaluate 
the finite volume functions contributing to  
the isovector magnetic form factor. 
These functions can be related to a more basic sum
that is ubiquitously encountered in these calculations
\begin{equation}
\cI_\a(\bm{B},\b^2) 
= 
\frac{1}{L^3} \sum_{\bm{k}} \frac{1}{[(\bm{k} + \bm{B})^2 + \b^2]^\a}
- 
\int \frac{d\bm{k}}{(2 \pi)^3}
\frac{1}{[(\bm{k} + \bm{B})^2 + \b^2]^\a}
,\end{equation}
which can be cast into an exponentially convergent form
involving elliptic theta functions~\cite{Sachrajda:2004mi}.
In the text, the function $\cK_2(m, \bm{B}, \D)$ is defined by
\begin{eqnarray}
\cK_2 (m , \bm{B},\D) 
&=& 
\frac{8 \pi^2}{L^3} \int_0^\infty d\l \sum_{\bm{k}} \frac{k_2 + B_2}{[(\bm{k} + \bm{B})^2 + \b_\D^2]^{3/2}}
,\end{eqnarray}
with $\b_\D^2 = \l^2 + 2 \D \l + m^2$. Using the expression for $\cI_{1/2}$ in terms
of elliptic theta functions, the integral over $\l$
can then be expressed in terms of the $\Erfc(x)$ function~\cite{Arndt:2004bg}. The final result appears as
the one-dimensional integral
\begin{equation}
\cK_2 (m , B \hat{y} ,\D) 
= 
- \frac{ \sqrt{\pi} L}{4}  \int_0^\infty d\tau \, 
\tau^{-5/2} e^{\tau (\D^2 - m^2)} \Erfc(\D \sqrt{\tau})
\vartheta'_3 \left( \frac{B L }{2} , e^{-\frac{L^2}{4 \tau}}\right)
\vartheta_3 \left( 0 , e^{-\frac{L^2}{4 \tau}}\right)^2
,\end{equation}
with $\vartheta_3(q,z)$ as the Jacobi elliptic theta function of the third kind.
Lastly the function $\cL_{33}(m,\bm{B},\D)$ is given by
\begin{eqnarray}
\cL_{33}(m, B \hat{y},\D) 
&=& 
8 \pi^2 \int_0^\infty  d\l
\left[
\frac{1}{L^3}
\sum_{\bm{k}}
\frac{(k_3)^2}{[(\bm{k} + \bm{B})^2 + \b_\D^2]^{5/2}}
- \int \frac{d \bm{k}}{(2 \pi)^3}
\frac{(k_3)^2}{[(\bm{k} + \bm{B})^2 + \b_\D^2]^{5/2}}
\right] 
\notag \\
&=& 
\frac{ \sqrt{\pi}}{3}  \int_0^\infty d\tau \, 
\tau^{-3/2} e^{\tau (\D^2 - m^2)} \Erfc(\D \sqrt{\tau})
\Bigg[ 
\vartheta_3 \left( 0 , e^{-\frac{L^2}{4 \tau}}\right)^2
\vartheta_3 \left( \frac{B L }{2} , e^{-\frac{L^2}{4 \tau}}\right)
\notag \\
&& \phantom{spacererer}
+ \frac{L^2}{8 \tau}
\vartheta''_3 \left( 0 , e^{-\frac{L^2}{4 \tau}}\right)
\vartheta_3 \left( 0 , e^{-\frac{L^2}{4 \tau}}\right)
\vartheta_3 \left( \frac{B L }{2} , e^{-\frac{L^2}{4 \tau}}\right)
- 1
\Bigg]
.\end{eqnarray}

\bibliography{hb}

\end{document}